\documentclass[aps,prl,twocolumn,amsmath,amssymb]{revtex4-1}
\newcommand{\kv}{{\mathbf k}} 
\newcommand{\rv}{{\mathbf r}} 
\newcommand{\psub}[1]{\psi_{\alpha_{#1}}}

\begin{document}

\noindent
{\bf Comment on ``Quantum Quasicrystals of Spin-Orbit-Coupled Dipolar Bosons''}\smallskip

In a recent Letter, Gopalakrishnan, Martin, and Demler~\cite[henceforth GMD]{Gopalakrishnan13} show that quasi-two-dimensional dipolar Bose gases, subject to a Rashba spin-orbit coupling, exhibit a variety of spatially ordered, or crystalline, ground states, including a pentagonal quasicrystal.
Indeed, as the authors say, realizing quasicrystalline condensates would provide new ways to explore the physics of quasicrystals, and in particular to study the quantum dynamics of their unique collective phason modes. Yet, the authors conclude that ``there are typically additional phasons in quantum-mechanical quasicrystals, when compared with their classical equivalents.'' In this Comment I review the notion of phason modes in quasicrystals, and explain why their number does not depend on whether they are classical or quantum.

Phonons and phasons are Goldstone modes that result from the fact that the crystalline phase breaks the continuous translational symmetry of the liquid phase, and therefore also of the hamiltonian or the free energy of the system. Any broken symmetry operation, by definition, takes a spontaneously chosen minimum free-energy state into a different minimum free-energy state. Consequently, the set of all broken symmetry operations provides a natural tool for exploring the space of all minimum free-energy states and for counting the number of independent phonon and phason modes~\cite{Sethna06}. See Ref.~\cite{LifshitzIJC11} for a detailed discussion of this in the classical context.

To be concrete, let us describe the crystalline state by a function $\psi_\alpha(\rv)$, possibly multicomponent---a real-valued scalar density or tensor field in the classical case, or a complex-valued wave function or spinor in the quantum mechanical case---whose Fourier expansion is given by
\begin{equation}\label{eq:DensityModes}
\psi_\alpha(\rv)=\sum_{\kv\in L}\psi_\alpha(\kv)e^{i\kv\cdot\rv},
\end{equation}
where the (reciprocal) lattice $L$ is a finitely generated $Z$-module. This means that $L$ can be expressed as the set of all integral linear combinations of a finite number $D$ of $d$-dimensional wave vectors~\footnote{In GMD, the dimension $d=2$, and the rank $D$ of an $M$-fold symmetric star of 2-dimensional wave vectors is given by the Euler totient function $\phi(M)=M\prod_{P_i}(1-1/P_i)$, where $P_i$ are the distinct prime factors of $M$.}. In the special case where the smallest possible $D$, called the \emph{rank} of the crystal, is equal to the physical dimension $d$, the crystal is periodic. More generally, for quasiperiodic crystals $D\geq d$, and one refers to quasiperiodic crystals that are aperiodic (with $D>d$) as \emph{quasicrystals} \cite{lifshitz03,*lifshitz07}.

A generic free energy, expressed in Fourier space as
\begin{eqnarray}\label{eq:F}
  \cal F &&= \sum_{j\leq n} \sum_{\alpha_1\ldots\alpha_n} \sum_{\kv_1\ldots\kv_n} A_j^{\alpha_1\ldots\alpha_n}(\kv_1,\ldots,\kv_n)\nonumber\\
&&\times\psub1(\kv_1)\ldots\psub{j}(\kv_{j})\psub{j+1}^\ast(-\kv_{j+1})\ldots\psub{n}^\ast(-\kv_{n}),
\end{eqnarray}
is restricted \emph{only} by the requirements imposed by the symmetry of the disordered liquid phase. In the absence of any external fields, the translation invariance of the disordered phase imposes the constraint that the free energy coefficients $A_j^{\alpha_1 \ldots \alpha_n}(\kv_1, \ldots, \kv_n)$ must vanish unless $\kv_1 + \ldots + \kv_{n} = 0$~\footnote{The constraints imposed by rotational invariance are immaterial here and treated elsewhere, for example in Ref.~\cite{LifshitzIJC11}.}. In the case of GMD the disordered phase has, in addition, an overall $U(1)$ symmetry $\psi_\alpha\to\psi_\alpha \exp[i\theta]$, which further constrains the free energy coefficients to vanish unless $n=2j$. 

A generic set of coefficients $A$ can vary freely with external parameters such as temperature and pressure subject to these restrictions and no others, while the field $\psi_\alpha$ adjusts itself accordingly to minimize $\cal F$. Thus, for two different fields $\psi_\alpha$ and $\psi'_\alpha$ to both minimize the same generic free-energy, and therefore be \emph{indistinguishable} by $\cal F$, they must agree independently on all the products in $\cal F$ with non-vanishing coefficients. In the absence of $U(1)$ symmetry, one can then show~\cite{Mermin92,RMPapp} using the identity of products of order 2 and 3 alone, that this condition for indistinguishability reduces to the requirement that 
\begin{equation}\label{eq:chi}
\psi_\alpha'(\kv)=e^{2\pi i\chi(\kv)}\psi_\alpha(\kv),
\end{equation} 
where $\chi(\kv)$ is a linear function to within an additive integer on the lattice $L$. Rokhsar, Wright, and Mermin~\cite{RWM88} called $\chi(\kv)$ a \emph{gauge function}, in analogy with gauge transformations in electrodynamics, because it encodes all the changes one can apply to the field $\psi_\alpha$ without affecting any of its physical observables. 

Any difference between classical crystals and quantum crystals \emph{with broken $U(1)$ symmetry} can arrise only from the additional constraint that $n=2j$. Nevertheless, one can still use the identity of products of the form $\psub1(\kv) \psub2^\ast(\kv)$ to establish Eq.~\eqref{eq:chi}; use products of order 4 to show that $\chi(-\kv) = -\chi(\kv)$; and products of order 6 to obtain $\chi(\kv_1 + \kv_2) = \chi(\kv_1) + \chi(\kv_2)$; with the latter equalities holding to within an overall constant coming from the broken $U(1)$ symmetry. Note that classical free energies may be restricted as well, by an additional symmetry of $\psi_\alpha\to - \psi_\alpha$, to contain products of only even order, where one similarly requires products of order 6 to establish the linearity of $\chi(\kv)$.

The linearity of gauge functions implies that $\chi(\kv)$ is uniquely determined by specifying its values on a chosen basis for $L$, consisting of $D$ linearly independent vectors over the integers. Thus, there are exactly $D$ independent symmetry operations in $\cal F$ that are broken, and therefore exactly $D$ Goldstone modes. One \emph{may} choose the basis so that $d$ of the Goldstone modes are the familiar phonon modes, which in the infinite wavelength limit reduce to rigid translations of the crystal. The remaining $D-d$ modes are phason modes, reducing in the infinite wavelength limit to relative phase shifts of the Fourier components of the crystal that leave some origin fixed.  

I should note that suppression of 6$^{th}$ or higher-order products at low density---whether in classical or in quantum crystals--- does not affect the conclusion here. Such terms might be irrelevant at the critical fixed point, but dangerously so~\cite{Amit82} in the sense that they still allow to distinguish between states in the ordered phase.

I am grateful to Michael Cross and Gil Refael for valuable discussions. This research is supported by the Israel Science Foundation through Grant No.~556/10.

\medskip\noindent
Ron Lifshitz\\
Condensed Matter Physics 149-33\\
California Institute of Technology\\
Pasadena, CA 91125, USA\\
On leave from:\\
School of Physics \& Astronomy\\
Tel Aviv University, Tel Aviv 66978, Israel

\medskip\noindent
Dated: December 4, 2013\\
PACS numbers: %
61.44.Br, 
64.70.Dv, 
67.85.-d, 
03.75.Mn 

\bibliography{phasons}

\end{document}